# Evaluation of Phosphorene as Anode Material for Na-ion Batteries from First Principles

*Vadym V. Kulish,[1,2]\* Oleksandr I. Malyi,[3] Clas Persson,[3,4] Ping Wu[1]\**

[1]Entropic Interface Group, Singapore University of Technology and Design, 20 Dover Drive, Singapore 138682, Singapore

[2]Institute of High Performance Computing, 1 Fusionopolis Way, #16-16 Connexis, Singapore 138632, Singapore

[3]Centre for Materials Science and Nanotechnology, and Department of Physics, University of Oslo, P.O. Box 1048 Blindern, NO-0316 Oslo, Norway

[4]Department of Materials Science and Engineering, Royal Institute of Technology, SE-100 44 Stockholm, Sweden

E-mails: vadym_kulish@sutd.edu.sg (V.V.K.), wuping@sutd.edu.sg (P.W.)

We systematically evaluate the prospects of a novel 2D nanomaterial, phosphorene, as anode for Na-ion batteries. Using first-principles calculations, we determine the Na adsorption energy, specific capacity and Na diffusion barriers on monolayer phosphorene. We examine the main trends in electronic structure and mechanical properties as a function of Na concentration. We find favorable Na-phosphorene interaction with theoretical capacity, exceeding those of alternative monolayer anodes for Na-ion batteries. We find that Na-phosphorene undergoes semiconductor-metal transition at high Na concentration. Our results show that Na diffusion on phosphorene is fast and anisotropic with the energy barrier of only 0.04 eV. Owing to the high capacity, good stability, excellent electrical conductivity and high Na mobility, monolayer phosphorene is a very promising anode material for Na-ion batteries. The calculated performance in terms of specific capacity and diffusion barrier is compared to other layered 2D electrode materials, such as graphene, $MoS_2$, polysilane, etc.

# INTRODUCTION

Layered black phosphorus (*"phosphorene"*) has emerged as a very promising 2D nanomaterial due to its direct band gap, high carrier mobility, ambipolar behavior and mechanical flexibility.[1, 2] The successful synthesis of single- and few-layer phosphorene by exfoliation method has been demonstrated recently.[2-7] Inside the phosphorene layer, each P atom is bonded with three neighboring P atoms, forming a puckered honeycomb structure similar to graphene. However, unlike zero-bandgap graphene, layered phosphorus has a direct band gap of 0.3 eV in the bulk form, which can increase up to 1.0-1.5 eV for an isolated monolayer.[3, 8] The direct band gap and ambipolar electric behavior of phosphorene make it very attractive for the optoelectronic devices operating in visible and near-infrared (NIR) regions.[9-12] Indeed, phosphorene-based transistors demonstrate high on/off ratios ($>10^5$) and good carrier mobilities (300-1000 $cm^2$ $V^{-1}$ $s^{-1}$).[2, 3, 8, 13-16] By combining phosphorene with $MoS_2$ or BN, it is possible to engineer 2D *p-n* junctions for photodetection and solar energy harvesting.[17, 18] The electronic and magnetic properties of phosphorene can be tuned by adatom adsorption,[19] oxygen passivation[20] or electric field.[21] Besides, unusual, non-flat and anisotropic structure of phosphorene leads to unique orientation-dependent electronic and optical properties, stress/strain response, etc.[21-27]

One of the emerging areas where phosphorene can find its potential application is the field of energy storage, particularly, sodium-ion (Na-ion) batteries. The Na-ion batteries are considered as very attractive low-cost technology for large-scale applications, such as smart electric grids that store clean renewable energy.[28-32] However, the identification of suitable negative electrodes is one of the main challenges for the future development of Na-ion batteries.[29] Recent studies show that alloy-type anode materials (*i.e.* Si, Sn, Sb, etc.) exhibit high specific capacities but suffer from poor reversibility due to the large volume expansion and slow Na kinetics.[33, 34] Anode materials following intercalation mechanism (*i.e.* $Na_3V_2(PO_4)_3$) demonstrate prolonged cycle life, but their capacity is limited.[35] Therefore, it is highly desirable to find anode materials which combine high capacity, good structural stability and fast Na diffusion. Phosphorus is one of the most promising anode materials due to its exceptionally high Na storage capacity, particularly, in the amorphous red phosphorus (RP) phase.[36-38] However, amorphous RP undergoes large volume changes upon Na insertion (~491%), resulting in large capacity fading with cycling.[36, 37] Moreover, amorphous RP is an insulator and has low electron conductivity, which seriously hinders the reversibility of Na insertion.[39, 40] More research is highly needed for the further development of P-based anodes, particularly on the crystal structure engineering, the nanosize effects, ionic and electronic conductivity optimization.

In this work, we propose a new candidate, monolayer phosphorene, as a promising Na-ion battery anode. Layered phosphorene is thermodynamically stable and has a higher electrical conductivity, than amorphous red phosphorus. Moreover, recent studies show that 2D materials demonstrate significant enhancements in battery performance as compared to their bulk counterparts.[41-43] Using first-principles calculations, we determine the Na adsorption energy, specific capacity and Na diffusion barrier on monolayer phosphorene. We investigate the main trends in electronic structure and mechanical properties of Na-phosphorene as a function of Na concentration. Our results demonstrate that phosphorene exhibits high theoretical capacity, good stability and fast Na diffusion rates. Owing to its large active surface area and much space available for Na diffusion, phosphorene is a promising candidate for Na-ion battery anodes.

## COMPUTATIONAL METHODS

First-principles calculations are performed within the density-functional theory (DFT) framework, as implemented in Quantum Espresso package.[44] The electron-ion interaction is described by a projector augmented wave (PAW) method[45] using Perdew-Burke-Ernzerhof (PBE) pseudopotentials.[46] The following valence electron configurations are used: P ($3s^2 3p^3$) and Na ($2s^2 2p^6 3s^1$). The kinetic-energy cutoff for valence electron wave functions is set as 35 Ry (476 eV). The Marzari-Vanderbilt cold smearing with a smearing factor of $\sigma = 0.01$ Ry (0.13 eV) is used in all calculations. The optimized structures are obtained by relaxing all atomic positions using the Broyden-Fletcher-Goldfarb-Shanno (BFGS) quasi-Newton algorithm until all forces are smaller than 0.01 eV/Å. The accuracy of our numerical procedure has been carefully tested. For phosphorene, we obtained P-P bond lengths of 2.22 Å for the horizontal and 2.26 Å for the bonds in other directions in good agreement with the recent studies.[19, 22, 47] The calculated band gap of monolayer phosphorene is 0.80 eV, consistent with other PBE studies[24] and underestimated by 25-50% from experiment and GW calculations.[3] This is expected with the PBE potential and does not affect our conclusions on Na adsorption energies and diffusion barriers. The calculated bulk lattice constant and cohesive energy of Na are 4.20 Å and 1.05 eV, respectively, close to the experimental values of 4.22 Å and 1.113 eV.[48]

Activation barriers for Na diffusion are calculated using the climbing-image nudged elastic band (CI-NEB) method.[49] It is an efficient technique for finding the minimum energy path between given initial and final positions. Here, the NEB calculations were performed with 7 images, and the initial guess of the migration pathway has been generated by linear interpolation between the initial and final points of the diffusion path. The geometry and energy of the images were then relaxed until the largest norm of the force orthogonal to the path was smaller than 0.01 eV/Å. This method has been used successfully to determine diffusion rates in various low-dimensional and bulk electrode materials.[43, 50-55] The amount of charge transfer between Na and phosphorene is estimated using the grid-based Bader charge method.[56]

To evaluate the effect of van der Waals (vdW) interactions on Na adsorption, we employ the semi-empirical correction scheme of Grimme (DFT-D2).[57] It has been shown that DFT-D2 approach provides the good accuracy in binding energies and diffusion barriers for Li-graphite systems.[58] We find that the contribution of vdW forces to Na adsorption energies on phosphorene ranges from 0.17 to 0.22 eV, as calculated for $NaP_{48}$ and $NaP_2$ systems, respectively. These values are consistent with the previous theoretical studies on Li adsorption on both pristine and defected graphene.[59, 60] Similar approach has been successfully used in theoretical studies of graphene[59] and monolayer $Ti_3C_2$ (MXene)[54] electrodes for Li-ion batteries.

## RESULTS AND DISCUSSIONS

### *Single Na atom insertion*

**Table 1.** Single Na atom insertion: adsorption energies ($E_a$), minimum Na-P distances ($d$), Bader charges on Na atoms ($Q_{Na}$)

| Site | $E_a$ (eV) (vs. Na-atom) ($\mu_{Na} = \mu_{Na}^{atom}$) | $E_a$ (eV) (vs. Na-bulk) ($\mu_{Na} = \mu_{Na}^{bulk}$) | $d(Na - P)$(Å) | $Q_{Na}(|e|)$ |
|---|---|---|---|---|
| H | -1.59 | -0.43 | 2.85 & 2x 2.93 | +0.85 |
| B | -1.51 | -0.36 | 2.82 | +0.85 |
| B1 | -1.16 | -0.003 | 2.87 | +0.82 |
| T | →H | →H | | |

We first examine what are the typical positions for Na insertion in phosphorene and study the nature of Na-phosphorene interaction. We consider four high-symmetry adsorption sites for single Na atom on phosphorene (Figure 1), namely: (1) above the center of P hexagon (*H* site), (2) above the mid-point of

P-P bond, which is located in the P-sublayer far from Na atom (*B* site), (3) above the mid-point of P-P bond, which is located in the P-sublayer near to Na atom (*B1* site), and (4) on top of P atom (*T* site). The Na adsorption energy ($E_a$) is defined as $E_a = E_{Na+P} - E_P - \mu_{Na}$, where $E_{Na+P}$ and $E_P$ are total energies of Na-adsorbed and pristine phosphorene, respectively, while $\mu_{Na}$ is chemical potential of sodium. By our definition, ($E_a < 0$) corresponds to exothermic reaction and attractive interaction. Table 1 summarizes the calculated adsorption energies and structural properties for all adsorption sites.

We find that the *H* site is the most energetically stable adsorption site. When Na atom is located at the *H* site, it has three nearest neighbors with Na-P distances of 2.85, 2.93 and 2.93 Å. All the neighboring P atoms move outwards from Na atom after relaxation, and the vertical P-P bond length in the $P_6$ ring underneath the Na atom is enlarged from 2.26 to 2.28 Å, indicating a slight weakening of the respective P-P bonds. When Na is at the *B* site, it has 2 nearest neighbors with Na-P distances of 2.82 Å. At the *B1* site, Na atom is again two-coordinated with the Na-P distances of 2.87 Å. The *B1* site is the least energetically favorable for Na adsorption (up to 0.43 eV difference), as also evidenced by the smallest charge transfer and the longest Na-P bond length (Table 1). The *T* site is not a stable position, since Na atom moves to *H* site after relaxation. These results suggest that Na atoms prefer adsorption sites with the maximum coordination number. The nature of Na-phosphorene binding has been examined by plotting the charge density difference ($\Delta\rho = \rho_{Na+P} - \rho_P - \rho_{Na}$) isosurfaces. Our analysis shows that there is a charge transfer from the Na atom towards the phosphorene, which is consistent with larger electronegativity of P relative to that of Na (2.19 and 0.93 on Pauling scale, respectively). The Bader charge analysis indicates that the Na atom carries a charge of 0.82-0.85 |e|, indicating that the bonding between Na and phosphorene has significant ionic character.

The appropriate choice of chemical potential of alkali atom (Li or Na) has been a subject of discussions recently.[59, 61, 62] Many previous theoretical studies have used the total energy of an isolated Li/Na atom as reference potential. However, it has been argued that in the evaluation of battery electrodes, the more suitable reference state should be the metallic bulk phase, instead of a neutral atom in the gaseous phase.[59, 61, 62] Moreover, the choice of Li/Na reference state may significantly affect the calculation results, as in the case of Li/Na-graphene. For instance, it was demonstrated that Na absorption on pristine graphene is not energetically possible relative to bulk metallic Na.[61] In our calculations, we employ both Na reference potentials ($\mu_{Na}^{atom}$ and $\mu_{Na}^{bulk}$). As shown in Table 1, under both conditions, the adsorption energy of Na on phosphorene is negative indicating favorable Na-phosphorene interaction.

The Na adsorption energy at the most stable site (*H*) (-1.59 eV) is much larger than the binding energy in $Na_2$ dimer (-0.76 eV) and Na bulk cohesive energy (-1.05 eV from our calculations). This suggests that Na would form 2D layers on the surface of phosphorene, and the clustering of Na atoms is not expected at the low Na concentration. This is distinctly different from the Na-graphene and Li-graphene, where the clustering of alkali atoms has been reported to hinder the anode performance.[62] Moreover, it has been recently argued that pristine monolayer graphene may not be suitable as anode for Li-ion and Na-ion batteries due to the weak alkali-graphene binding.[62] Alkali-graphene interaction is quite weak even at low concentrations, unless the reactive edges, defects or doping are introduced. In contrast, Na-phosphorene binding is energetically favorable owing to the higher reactivity of $sp^3$-like hybridization of P atoms.

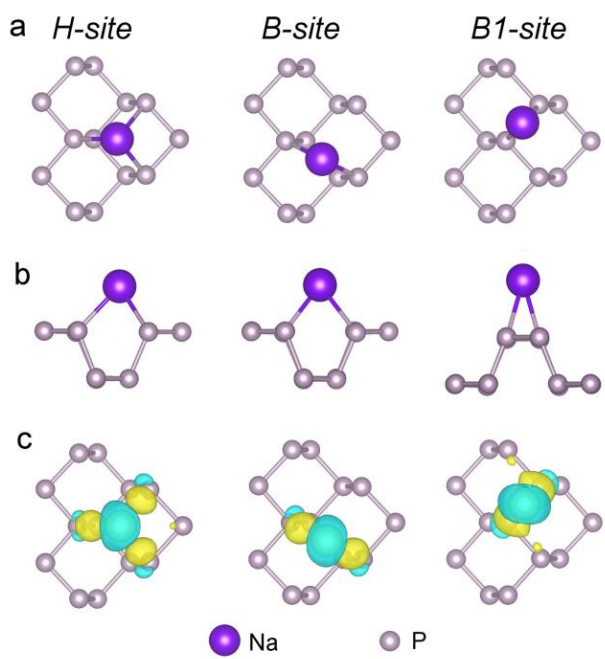

**Figure 1.** (a) Top and (b) side views on the stable Na adsorption sites (*H*, *B* and *B1*) and (c) corresponding charge density difference isosurfaces. The blue and yellow colors represent charge depletion and charge accumulation regions, respectively.

### *Multiple Na insertion*

We further study Na insertion at different concentrations by gradually increasing the amount of Na atoms in the system. For each Na concentration, we consider various possible initial Na orderings. In all configurations, Na atoms are initially placed in either *H* or *B* sites as far away from each other as possible in order to minimize their self-interaction. All structures are then fully optimized. The Na ordering with the lowest binding energy for each concentration is chosen for further analysis.

We consider both single-side and double-side adsorption of Na atoms on phosphorene. Figure 2 shows the most stable atomic configurations for single-side $Na_xP$ compounds ($x$ = 0.0625, 0.125, 0.25, 0.5), corresponding to $NaP_{16}$, $NaP_8$, $NaP_4$, $NaP_2$, respectively. We find that in the relaxed structures, all of the Na atoms remain isolated and keep their positions above the centers of the hexagons ($H$ sites). In the structure with the highest Na coverage, Na atoms occupy all hollow sites with the Na-Na distance of 3.00 Å. In comparison, the separation between the adjacent $H$ sites in graphene is only 2.46 Å. Therefore, larger Na concentrations may be expected in phosphorene than in graphene due to much reduced Na-Na repulsion.

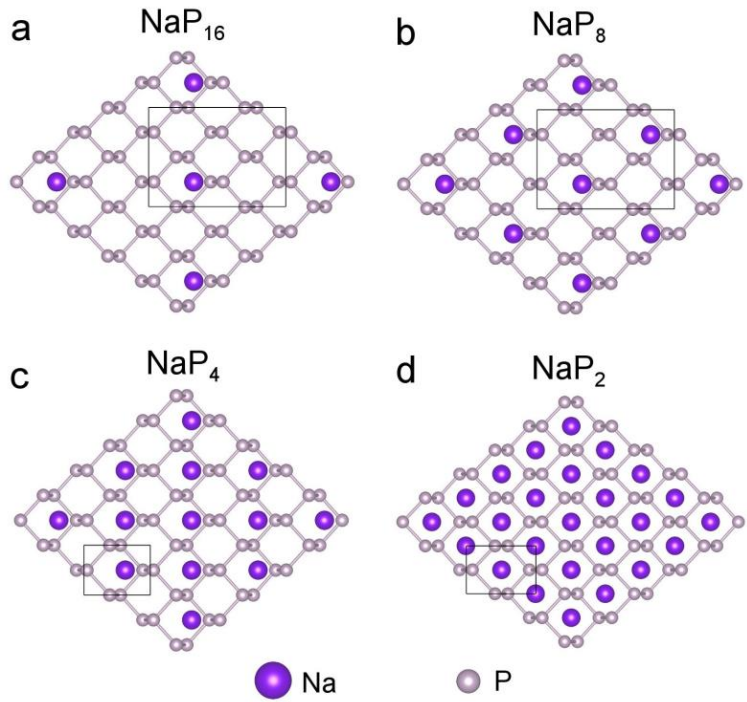

**Figure 2.** Structures of the Na-adsorbed phosphorene at difference Na coverages for single-side adsorption. The Na concentration ($x$) is equal to (a) 0.0625, (b) 0.125, (c) 0.25 and (d) 0.5. The rectangles represent the supercells for each case.

The stability of the studied $Na_xP$ phases can be evaluated from their formation energy

$$E_{form}(x) = \frac{1}{x}(E(Na_xP) - E(P) - x\mu_{Na})$$

where $E(Na_xP)$ and $E(P)$ are total energies of Na-adsorbed and pristine phosphorene, respectively, $\mu_{Na}$ is chemical potential of sodium, while $x$ is concentration.

Figure 3a shows a variation in the formation energy for single- and double-side adsorption as a function of Na concentration. The formation energy is negative at low Na content, suggesting that Na-P nanostructures are stable. As the Na concentration increases, $E_{form}$ becomes smaller. The reduction in $E_{form}$ is attributed to the two main factors. The first is weakened electrostatic attraction between phosphorene host and Na cations. Figure 3b shows the calculated partial charge on Na atoms (0.85, 0.56 and 0.34 |e| in $NaP_{48}$, $NaP_4$ and $NaP_2$, respectively). The charge transfer from Na to phosphorene is reduced as much as twice at high Na content. The second is the enhanced Na-Na repulsion at high Na concentrations. The overall Na adsorption energy results from the competition between these two factors: repulsive Na-Na forces and the attraction between Na and phosphorene host. For the single-side adsorption, the formation energy $E_{form}$ changes from negative to positive at about $x = 0.5$ (see Figure 3a). The calculated $E_{form}$ of $NaP_2$ phase for the single-side adsorption is 0.022 eV. Since this value is even smaller than the $k_BT$ at room temperature (0.026 eV), we consider $NaP_2$ as the maximum composition for single-side adsorption.

We find that for any Na concentration, double-side adsorption is always more stable than single-side, which is attributed to the minimization of Na-Na repulsion in the former case. In case of double-side adsorption, Na atoms prefer to adsorb on the different sides of the same $P_6$ ring. We find that such configuration is more stable than far-away configuration by 0.05 eV per Na atom, since it minimizes the deformations of phosphorene layer upon Na insertion. For the double-side adsorption, $E_{form}$ changes from negative to positive at $x \approx 1$. This indicates that NaP phase ($x = 1$) with $E_{form} = -0.06$ eV is the maximum achievable composition in monolayer Na-phosphorene. Note that the inclusion of vdW dispersion term has a significant impact on the determination of maximum $Na_xP$ composition. We find that the dispersion correction has improved the thermodynamic stability of NaP phase by about 0.15 eV. Consequently, NaP is the highest accessible Na composition according to DFT-D2, while $NaP_2$ is the maximum phase according to standard DFT, as shown in Figure 3a.

From the obtained maximum $Na_xP$ compositions, we can evaluate the theoretical specific capacity of monolayer phosphorene. Our calculated theoretical specific capacities are 865.30 (NaP) and 432.65 ($NaP_2$) mAh/g for the double- and single-side Na adsorption, respectively. The high capacity of phosphorene is attributed to the effective reduction of high electrostatic repulsion between Na atoms. In phosphorene, the separation between adjacent H sites is 3.00 Å, which is close to the Na-Na bond length of 3.079 Å in Na$_2$ dimer. In contrast, the separation between the adjacent $H$ sites in graphene is only 2.46 Å. In comparison to other two-dimensional anode materials for Na-ion batteries, monolayer phosphorene demonstrates larger theoretical capacity than MoS$_2$ (146 mAh/g),[63] Ti$_3$C$_2$ MXenes (352 mAh/g)[64] and boron-doped graphene (423-762 mAh/g).[61] Moreover, the capacity values, calculated by both DFT-D2 and DFT methods, are larger than that of the commercially used graphite anode (372 mAh/g) in Li-ion batteries.

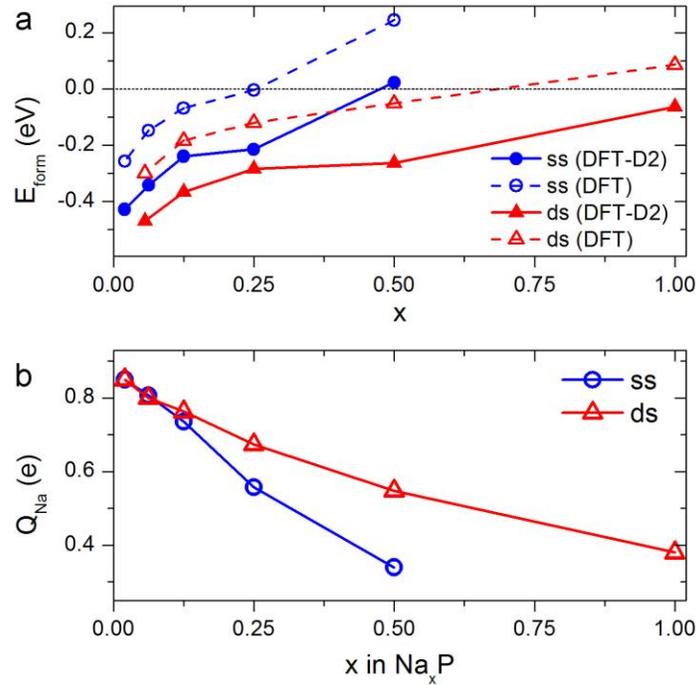

**Figure 3.** (a) Formation energy, (b) Bader charge on each Na atom as a function of Na concentration for single *(ss)* and double-side *(ds)* Na adsorptions.

### *Effect of sodiation on the mechanical properties of Na-phosphorene*

Sodiation and lithiation are known to produce large changes in the mechanical properties of electrode materials, especially alloys.[65] Moreover, the deformations and stress due to sodiation can cause the electrode to fracture or change its morphology, eventually leading to large capacity fading.[66, 67] To address the effects of Na insertion on the mechanical properties of phosphorene, we calculate its in-plane stiffness ($C$) – the analogue of Young's modulus for 2D materials. This can be done by applying a series of small deformations on the material and examining its mechanical response. The applied strain is defined as $\varepsilon = \frac{a-a_0}{a_0}$, where $a$ and $a_0$ are the lattice constants of strained and relaxed Na-P compound, respectively. The strain energy is defined as $E_s = E(\varepsilon) - E(0)$; namely, the difference between the total energies at a given strain $\varepsilon$ and at the zero strain. Starting with the fully relaxed structure of Na-phosphorene, its lattice constants are imposed to small strains $\varepsilon_x$ and $\varepsilon_y$ within the harmonic range of ±0.02 (Figure 4a). For each ($\varepsilon_x, \varepsilon_y$) value, we fully optimize the Na-P structure and calculate the strain energy $E_s$. In the harmonic regime, the strain energy can be fitted to a two-dimensional quadratic polynomial expressed by

$$E_s(\varepsilon_x, \varepsilon_y) = a_1 \varepsilon_x^2 + a_2 \varepsilon_y^2 + a_3 \varepsilon_x \varepsilon_y$$

Through fitting of strain-energy surface, we can derive the in-plane stiffness ($C$) of Na-phosphorene as follows[68]

$$C_x = \frac{4a_1 a_2 - a_3^2}{2a_2 A_0} \text{ and } C_y = \frac{4a_1 a_2 - a_3^2}{2a_1 A_0}.$$

The calculated in-plane stiffness for pristine monolayer phosphorene is equal to $C_x$=30 and $C_y$=94 N/m, in good agreement with the available theoretical studies.[69] Interestingly, the in-plane stiffness of phosphorene in the zigzag (*y*) direction is >3 times larger than in the armchair (*x*) direction, reflecting the anisotropic nature of the material. The low in-plane stiffness and excellent mechanical flexibility of phosphorene suggest its ability to withstand large strains without breaking which is beneficial for the application in Na-ion batteries where sodiation typically induces large stresses and deformations.

Figure 4b shows the variation of the in-plane stiffness of Na-phosphorene compounds as a function of Na concentration. $C_x$ decreases from 30 to 17 N/m (up to 43%) with increasing Na content demonstrating elastic softening of phosphorene host. Meanwhile, $C_y$ changes by less than 4% (from 94 to 88 N/m) upon Na insertion. These results demonstrate the highly anisotropic elastic properties of monolayer phosphorene with the host matrix being much more rigid in the zigzag (*y*) direction. The softening effect is attributed to the reduction in P-P bond strength after sodiation. It is well correlated with an increase in P-P bond length at higher Na concentrations. The changes in elastic properties of Na-phosphorene are considerably small. In contrast, sodiation of Na$_x$M (M = Sn, Pb, Si, Ge) bulk alloys leads up 75% deterioration of their elastic moduli.[70]

The crystal deformation and expansion caused by sodiation are relatively small. Since phosphorene is 2D nanomaterial, we have analyzed the variation of specific area $A = a_x \cdot a_y$ after Na insertion. The calculated changes defined as $A/A_0$ are 1.5, 3 and 5% for $NaP_8$, $NaP_4$ and $NaP_2$ phases with single-side Na adsorption, respectively. For the double-side Na adsorption, the area changes are slightly larger, namely: 1, 2, 9 and 16% for $NaP_8$, $NaP_4$, $NaP_2$ and fully-sodiated NaP compounds, respectively. These values are much smaller than the volume expansion of 491% in bulk red phosphorus anodes.[37] Interestingly, the strain caused by sodiation is accommodated by the anisotropic expansion of phosphorene layer in $x$ direction. We find that expansion in armchair ($x$) direction is 3 times larger than in zigzag ($y$). The obtained results show high mechanical stability and integrity of phosphorene anode which is very important for the prolonged cycle life of the battery.

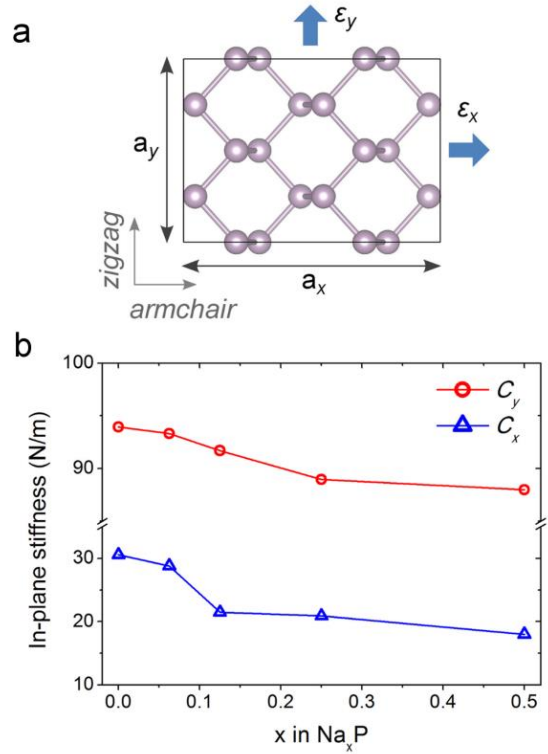

**Figure 4.** Mechanical properties of $Na_xP$. (a) Top view of the typical phosphorene supercell used for the calculations of elastic constants. $a_x$ and $a_y$ are the lattice constants of the supercell in $x$- and $y$-directions, respectively. $\varepsilon_x$ and $\varepsilon_y$ are strains in $x$ and $y$ directions, respectively (b) In-plane stiffness ($C$) as a function of Na concentration

*Electronic properties of Na-phosphorene*

Next, we analyze the changes in electronic structure of monolayer phosphorene upon Na insertion. Figure 5 shows the concentration-dependent total density of states (DOS) for different Na-phosphorene compounds. At low Na concentration, the total DOS of Na-P is similar to that of pristine phosphorene host. Na states are mainly located in the conduction band region. Adsorption of Na atoms induces the charge transfer, which shifts the Fermi level into the conduction band. Increasing the number of Na atoms also increases the amount of total charge transfer. Therefore, the Fermi level is shifted even further, as shown in Figure 5. At moderate and large Na concentrations, the DOS of Na-phosphorene exhibits significant changes. In particular, the bandgap of Na-P compound gradually diminishes and eventually disappears. We find that nanostructures with large Na content are metals. Therefore, we observe a semiconductor-metal transition with increasing Na concentration which is beneficial for the prospective application as Na-ion battery anode.

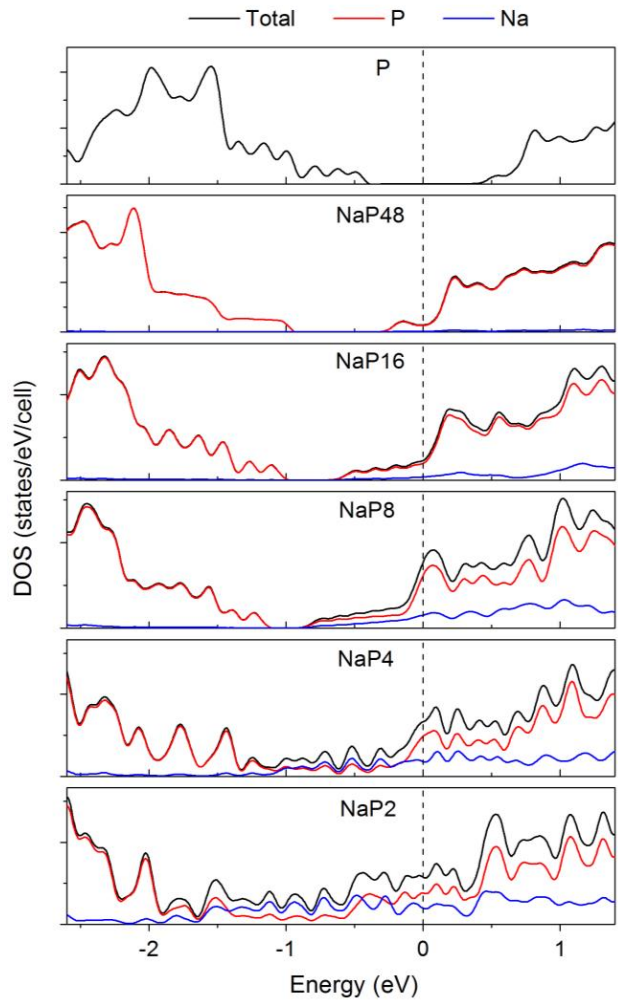

**Figure 5.** Concentration-dependent DOS for Na-phosphorene. The zero energy (vertical dashed line) is set to the Fermi level

*Na diffusion*

The rate capability of electrode material depends on the kinetics of electron transport and intrinsic Na-ion diffusion. Latest studies demonstrate good carrier mobilities of ~300-1000 cm$^2$ V$^{-1}$ s$^{-1}$ in few-layer phosphorene.[2, 3] Moreover, we have shown that at high Na concentrations, the bandgap of Na-phosphorene vanishes and the system becomes metallic. Hence, the electron transport in phosphorene anode is expected to be sufficiently fast.

Examination of Na mobility and diffusion pathways in the electrode material is of crucial importance. Simulations of such process can greatly enhance our understanding of ion diffusion pathways by evaluating the activation energies for various possible mechanisms at the atomic level. Fast Na diffusion is desirable for a high charge/discharge rates in practical Na-ion batteries. As shown above, the *H* site is the most stable position for Na atom on the phosphorene surface. The diffusion of Na in phosphorene occurs in the form of jumps between neighboring *H* sites, while bridge positions (*B* and *B1*) serve as transition states. When a single Na atom is placed on the phosphorene surface, there are three possible directions for Na diffusion, namely: *H→B→H*, *H→B1→H* and *H→T→H*. The calculated diffusion pathways and the corresponding energy profiles are shown in Figure 6.

We have found that *H→B→H* is the fastest Na diffusion pathway on phosphorene surface. The calculated energy barrier for Na diffusion on phosphorene via *H→B→H* path is only 0.04 eV. At the saddle point, Na changes its state from three-coordinated to two-coordinated leading to an energy barrier. Meanwhile, the smaller charge transfer at *B1* site and larger deformation of phosphorene monolayer destabilize the *B1* transition state as compared to *B* site. Consequently, the activation energy for *H→B1→H* diffusion pathway is increased up to 0.36 eV. We have also studied the possibility of Na diffusion via *H→T→H* path and found that the corresponding energy barrier is about 0.38 eV.

Overall, our results demonstrate that Na diffusion on phosphorene is quite anisotropic. The diffusion anisotropy is caused by a non-flat, puckered structure of phosphorene providing fast Na diffusion channels. The calculated ratio between energy barriers in the orthogonal directions is equal to $E_{zigzag}/E_{armchair} \approx 1/8$. This behavior is quite interesting and unusual and has not been observed in any other 2D nanomaterials (*i.e.* graphene, MoS$_2$, silicene, etc.).

The calculated activation energy for Na diffusion on phosphorene (0.04 eV) is smaller than on graphene (0.10eV), boron-doped graphene (0.16-0.22 eV),[61] layered polysilane (0.41 eV),[50] monolayer H-passivated silicene (0.12 eV)[50] and MoS$_2$ (0.28 eV)[63]. Moreover, the calculated value is smaller than the energy barrier for Li diffusion in commercially used graphite (0.22 eV)[71] and high-capacity bulk silicon anode (0.57 eV).[43, 53, 72, 73] Due to large surface area of phosphorene and low activation energy for Na surface diffusion, we can expect a significant improvement of Na diffusion rates in phosphorene as potential anode materials for Na-ion batteries.

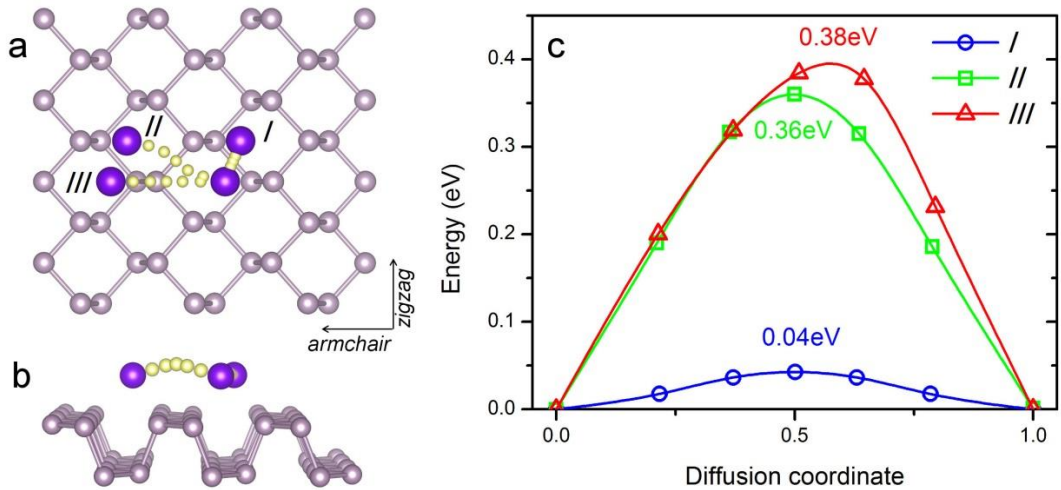

**Figure 6.** Na diffusion on monolayer phosphorene. (a) Top and (b) side views on Na diffusion pathways: (*I*) H→B→H, (*II*) H→B1→H and (*III*) H→T→H. (c) Corresponding energy barriers for Na diffusion on phosphorene. Yellow circles show intermediate atomic positions in the diffusion pathways

## Conclusions

In summary, we have systematically evaluated the prospects of a novel monolayer material, phosphorene, as anode for Na-ion batteries. Using first-principles calculations, we determined the Na adsorption energy, theoretical specific capacity and diffusion barriers on monolayer phosphorene. We investigated the main trends in electronic structure and mechanical properties as a function of Na concentration. We found that the adsorption energy of single Na atom on phosphorene is negative indicating favorable Na-phosphorene interaction. Moreover, the Na adsorption energy on phosphorene is much larger than the binding energy in $Na_2$ dimer and Na bulk cohesive energy suggesting that the clustering of Na atoms is not expected at the low Na concentration. Monolayer phosphorene demonstrates exceptionally high theoretical specific capacity, being capable to adsorb Na atoms up to the compositions of NaP (865 mAh/g) and $NaP_2$ (433 mAh/g) as predicted by DFT-D2 and DFT methods, respectively. Phosphorene exhibits high mechanical stability and integrity upon Na insertion.

We found that Na-phosphorene is metallic at high Na content, which is a significant advantage in electronic conductivity over amorphous red phosphorus phase. Our results show that Na diffusion on phosphorene is very fast with the energy barrier of only 0.04 eV. Interestingly, we have found the Na diffusion on the surface phosphorene is anisotropic (the energy barrier in zigzag direction is 8 times smaller, than in the other directions). Importantly, the diffusion barrier in both directions is satisfactory from the viewpoint of battery application. Considering the beneficial properties of phosphorene, such as high stability, high theoretical capacity, excellent electrical conductivity, large active surface area and fast Na diffusion, monolayer phosphorene is a very promising anode material for Na-ion batteries. With the availability of synthesized phosphorene, we hope that our results will inspire future experimental and theoretical studies.

**Acknowledgements**

This work was supported by a Ministry of Education of Singapore (MOE Tier 2 grant 2012-T2-1-097). The authors appreciate computational support received from the Institute of High Performance Computing of A*STAR of Singapore. The authors acknowledge the support from the Research Council of Norway (contract No. 221469) and access to HPC resources at NSC through Swedish SNIC/SNAC and at USIT through Norwegian NORTUR.